\documentclass[preprintnumbers, prd, onecolumn, showpacs, floatfix, preprintnumbers, superscriptaddress, nofootinbib]{revtex4}

\usepackage{graphicx}
\usepackage{epsfig}
\usepackage{bm}
\usepackage{amssymb}
\usepackage{float}
\usepackage{amsmath}
\usepackage{subfigure}
\usepackage{dcolumn}
\usepackage{epstopdf}
\usepackage{cancel}
\usepackage[colorlinks]{hyperref}
\usepackage[usenames,dvipsnames]{color}
\hypersetup{
     breaklinks=true,
    pdfstartview={FitH},    % fits the width of the page to the window
    colorlinks=true,       % false: boxed links; true: colored links
    linkcolor=blue,          % color of internal links
    citecolor=red,        % color of links to bibliography
    filecolor=magenta,      % color of file links
    urlcolor=blue,           % color of external links
    anchorcolor=green,      % Color for anchor text
    linktocpage=true
}

\newcommand{\Mpl}{M_{\textrm{Pl}}}
\renewcommand{\(}{\left(}
\renewcommand{\)}{\right)}

\def\lam{\lambda}

\def\S{\mathcal{S}}

\def\doi{http://doi.org}
\def\arxiv{http://arxiv.org/abs}

\def\e{\mathrm{e}}
\def\r{\mathrm{r}}

\def\m{\mathrm{m}}

\begin{document}

 \title{ Observational constraints on successful model of quintessential Inflation }

\author{Chao-Qiang Geng}
\email{geng@phys.nthu.edu.tw}
\affiliation{Chongqing University of Posts \& Telecommunications, Chongqing, 400065, China}
\affiliation{National Center for Theoretical Sciences, Hsinchu, Taiwan 300}
\affiliation{Department of Physics, National Tsing Hua University, Hsinchu, Taiwan 300}

\author{Chung-Chi Lee}
\email{lee.chungchi16@gmail.com}
\affiliation{DAMTP, Centre for Mathematical Sciences, University of Cambridge, Wilberforce Road, Cambridge CB3 0WA, UK}

\author{M.~Sami}
\email{sami@iucaa.ernet.in} \affiliation{Centre for Theoretical Physics, Jamia Millia Islamia, New Delhi-110025, India}

\author{Emmanuel~N.~Saridakis}
\email{Emmanuel\_Saridakis@baylor.edu}
\affiliation{Physics Division, National Technical University of Athens, 15780 Zografou Campus, Athens, Greece}
\affiliation{CASPER, Physics Department, Baylor University, Waco, TX 76798-7310, USA}

\author{Alexei A. Starobinsky}
\email{alstar@landau.ac.ru} \affiliation{ L.~D.~Landau Institute for Theoretical Physics RAS, Moscow 119334, Russian Federation}
\affiliation{Bogolyubov Laboratory of Theoretical Physics, Joint Institute for Nuclear Research,  \\ Dubna 141980, Russian Federation}

\begin{abstract}

We study quintessential inflation using a generalized exponential potential $V(\phi)\propto \exp(-\lambda \phi^n/\Mpl^n),~n>1$, the model admits slow-roll inflation at early times and leads to close-to-scaling behaviour in the post inflationary era with an exit to dark energy at late times. We 
present detailed investigations of the inflationary stage in the light of the Planck 2015 results, study post-inflationary dynamics and analytically 
confirm the existence of an approximately scaling solution. Additionally, assuming that standard massive neutrinos are non-minimally coupled, makes the field $\phi$ dominant once again at late times giving rise to present accelerated expansion of the 
Universe. We derive observational constraints on the field and time-dependent neutrino masses. In particular, for $n=6~(8)$, the 
parameter $\lambda$ is constrained to be,
$\log \lambda > -7.29 ~(-11.7)$; 
 the model produces the spectral index of the power spectrum of primordial scalar (matter density) perturbations as 
$ n_s = 0.959 \pm 0.001~ (0.961 \pm 0.001)$ and tiny tensor-to-scalar ratio,  $r<1.72 \times 10^{-2}~ (2.32 \times 10^{-2})$ 
 respectively. Consequently, the upper bound on possible values of the sum of neutrino masses $\Sigma m_{\nu} \lesssim 2.5$~eV  
significantly enhances compared to that in the standard $\Lambda$CDM model.
\end{abstract}

\pacs{98.80.-k, 95.36.+x, 98.80.Cq}

\maketitle

\section{Introduction}

\label{sec:introduction}

It is remarkable that according to the standard cosmological paradigm,  both early and late-time phases of the Universe history require accelerated expansion. Inflation \cite{Starobinsky:1979ty,Starobinsky:1980te,Guth:1980zm,Linde82,Linde:1983gd} naturally 
provides necessary initial conditions, including small perturbations, for the subsequent isotropic hot Big Bang (radiation dominated) stage at 
early times, whereas the qualitatively similar late-time cosmic acceleration ~\cite{SS00,Peebles:2002gy,Copeland:2006wr, Sahni:2006pa, Adam:2015wua, Aghanim:2015xee} beautifully resolves the age crisis and fits the SNIa luminosity distance and baryon acoustic oscillations 
(BAO) data. Hence, the radiation dominated (RD) and matter dominated (MD) stages of the Universe evolution are sandwiched between 
two phases of accelerated expansion. In this perspective, it sounds quite reasonable to imagine a common origin for both the phases, in other 
words, a model of inflation which could also give rise to late-time acceleration. 

The idea of unification of inflation and late-time acceleration (even
before an observational discovery of the latter) was first briefly mentioned in \cite{Spok93}, and then further developed in~\cite{Peebles:1999fz} where the name {\it quintessential inflation} was introduced, too, see also \cite{Peloso:1999dm,Dimopoulos:2000md,Dimopoulos:2001ix,Giovannini:2003jw,
Sami:2004ic,Rosenfeld:2005mt,Hossain:2014xha,Hossain:2014coa}.
Later, it implemented in the framework of braneworld cosmology in~\cite{Copeland:2000hn,Sahni:2001qp,Majumdar:2001mm,Sami:2004xk} (for a review see Ref.\cite{Hossain:2014zma}, see also Ref.\cite{Dimopoulos:2017zvq} on the related theme).

It is indeed quite challenging to describe both phases of acceleration using a single scalar field minimally couple to gravity, without affecting 
the thermal history of the universe which has been verified to a good accuracy. Let us spell out the broad requirements on the field, if it is
to unify both the phenomena. In order to facilitate slow-roll, the field potential should exhibit shallow behaviour at early times, followed by a steep 
region for most of the universe history, turning shallow once again at late times. Since the inflaton scalar field survives till late  in this picture,  
conventional mechanisms of creation and heating of matter after the end of inflation by the inflaton field itself, either perturbative ones~\cite{ASTW82,DL82,Abbott:1982hn} or using the non-perturbative broad parametric resonance (preheating)~\cite{Kofman:1994rk,Kofman:1997yn}, are 
not operative, while those based on the effect of gravitational particle creation~\cite{Parker:1968mv,ZeldovichStaro1,ZeldovichStaro2,Starobinsky:1980te,Ford87} may appear to be insufficient, too (though this depends on 
a value of the dimensionless model parameter $\lambda$ and requires special consideration~\footnote{Since gravitons are created by the 
same effect, too, in this case the problem of gravitational wave overproduction may arise.}). A good alternative may be provided by 
preheating based upon instant particle production ~\cite{Felder:1998vq}, which can successfully address the problem of relic gravitational 
waves generated during inflation~\cite{Starobinsky:1979ty,Sahni:1990tx,Souradeep:1992sm,Giovannini:1999qj} and after 
it~\cite{Grishchuk:1974ny}.

After inflation ends, the scalar field enters  the kinetic regime~\cite{Spok93,JP98,Sahni:2001qp,Sami:2004xk}, and overshoots the 
background and freezes on its potential. During evolution, the background energy density becomes comparable to the field energy density, 
thereafter the field evolution crucially depends upon the degree of steepness of its potential. In case of a steep exponential potential~\cite{Lucchin:1984yf,Ratra:1989uz,Copeland:2000hn,Sahni:2001qp,Sami:2002fs,Hossain:2014xha}, the field would track the background 
being sub-dominant, in the so-called scaling regime  \cite{Steinhardt:1999nw}. If the potential is less steeper than the exponential function, the 
field energy density would move towards the background, as it happens in case of inverse power-law potentials. However, in the opposite case, 
the field would move away from the background, overshoot it and freeze on its potential, and this latter behavior would keep repeating.

A new phenomenon occurs if we consider the class of potentials $V(\phi)\sim \exp(-\lambda \phi^n/\Mpl^n)$~\cite{Geng:2015fla}. First, one 
can realize successful inflation in this model for suitable values of $\lambda$ and $n$. Second, in the post-inflationary evolution, the model 
exhibits an interesting behavior. Indeed, since $\Gamma=V_{\phi\phi}V/V_{\phi}^2\to 1$ for large values of $\phi$, the model can give rise to an
approximately scaling solution at late stages, before which the field exhibits the aforementioned behavior. With the steep exponential potential, 
one could invoke high energy brane corrections~\cite{Randall:1999ee,Randall:1999vf,Sahni:2001qp,Sami:2004xk,Copeland:2000hn}
to facilitate slow-roll. Assuming the required late-time features in the potential, this construction could give rise to viable
post-inflationary evolution.

Although the simple exponential potential does not comply with the observational constraints related to inflation, the generalized exponential gives 
rise to slow-roll for small values of $\phi$~\cite{Sami:2004xk,Sami:2004ic,Hossain:2014xha,Hossain:2014coa,Hossain:2014zma}. In this case 
we have one extra parameter at hand that might allow us to satisfy all observational constraints for inflation. At the same time, the model enjoys 
all the benefits of the steep exponential case, as it effectively mimics it at late stages.

Because of its approximately scaling behaviour, the field $\phi$ by itself may not provide the late time transition to the present 
accelerated expansion of the Universe. To achieve this, some new element has to be added to the model. One possibility is to assume a 
non-minimal coupling of the field $\phi$ to all matter~\cite{Amendola:1999er,Gumjudpai:2005ry}. In particular, a large coupling constant gives 
rise to a minimum of the effective potential which can trap the field producing an attractor of the dynamical system that mimics the 
cosmological-constant-like behavior. However, the latter is undesirable as the said regime can be reached soon after the transition from 
radiation to matter dominance in the early Universe. Thus, to have a sufficiently long MD stage required by observations, it is 
absolutely essential to leave the main part of non-relativistic matter (including cold dark matter and baryons) intact.  On the other hand, a 
non-minimal coupling of  $\phi$ to neutrinos only, that makes neutrino masses time-dependent, is well possible, and it can safely trigger 
a late-time transition to the present accelerated expansion of the Universe leaving the 
duration of the MD stage practically unchanged~\cite{Fardon:2003eh, Bi:2003yr, Hung:2003jb, Peccei:2004sz, Bi:2004ns, Afshordi:2005ym, Brookfield:2005bz, Amendola:2007yx, Wetterich:2007kr, Pettorino:2010bv, Ayaita:2014una}.

In this case, the coupling forms  dynamically as massive neutrinos become non-relativistic at late times. Such a coupling seems to be a natural 
device for triggering the transition from the scaling regime to the late-time acceleration.

In this paper, we explore aspects of quintessential inflation using the new generalized exponential potential \ref{eq:potential} for the 
inflaton-quintessence scalar field $\phi$, both without and with its coupling to massive neutrinos according to the formula~\ref{eq:veff}. In 
Section~\ref{sec:2} we describe the model setup, and in Section  \ref{sec:inflation} we present analytical and numerical details of the inflationary 
stage in this model. Section~\ref{sec:post_inf} includes the major part of our work and is devoted to investigations of observational constraints 
on the sum of neutrino masses. This section additionally includes an analytical proof of the existence of a scaling solution in the model under 
consideration. Our results are summarized in Section~\ref{sec:conclusion}.

\section{Non-minimally coupled massive neutrino matter to dark energy}

\label{sec:2}

 In th We are are interested in a scenario which could give rise
to successful inflation at early times followed by a viable post inflationary evolution with a possibility of exit to acceleration at late stages. In order to accomplish the underlying idea, we  
 consider the following action (we use the metric signature, ($+,-,-,-$)),
% \cite{Wetterich:2013aca}
\begin{eqnarray}
\mathcal{S} = \int d^4x \sqrt{-g}\bigg[-\frac{\Mpl^2}{2}R+
\frac{1}{2}(\nabla \phi)^2-V(\phi)
\bigg]+\S_\m+\S_\r+\S_\nu\(\phi, \Psi\) \, ,
\label{eq:action}
\end{eqnarray}
where $S_{m, r, \nu}$ correspond to the actions of matter, radiation
and neutrino matter, respectively. We assume here that
%As discussed in the introduction, the standard matter is not coupled to field whereas
only massive neutrino
matter is non-minimally coupled to the scalar field $\phi$ with the Lagrangian,
\begin{eqnarray}
\mathcal{L}_{\nu} = i \bar{\Psi} \gamma^{\lambda} \partial_{\lambda} \Psi - m_{\nu}
\bar{\Psi}  \Psi = i \bar{\Psi} \gamma^{\lambda} \partial_{\lambda} \Psi - m_{\nu,0}
e^{\beta \phi} \bar{\Psi} \Psi \,.
\label{eq:coupling}
\end{eqnarray}
Varying the action (\ref{eq:action}) with respect to the metric and keeping in mind the
Friedmann-Lema\^itre-Robertson-Walker (FLRW) spatially flat background, 
we obtain the evolution equations
\begin{eqnarray}
&&3H^2\Mpl^2 = \frac{1}{2}\dot\phi^2+V(\phi)+\rho_\m+\rho_\r+\rho_\nu \, ,
\label{eq:FR1} \\
&&\(2\dot H+3H^2\)\Mpl^2 = -\frac{1}{2}\dot\phi^2+V(\phi)-p_m-p_\r-p_\nu \,,
\label{eq:FR2}
\end{eqnarray}
where $\rho_i$ ($p_i$) are the energy densities (pressures) of the
corresponding sectors. Additionally, varying the action
(\ref{eq:action}) with respect to the scalar field $\phi$, we derive its
equation of motion, which reads as
\begin{equation}
\label{eq:phieq}
\ddot\phi+3H\dot\phi+\frac{d V_{\rm {eff}}}{d \phi} =0 \,,
\end{equation}
where $V_{\rm {eff}}$ is the effective potential such that $dV_{\rm {eff}}/d
\phi= dV(\phi)/d\phi + \beta(\rho_\nu-3p_\nu)/\Mpl$. Consequently,
the evolution equation for neutrino matter becomes
\begin{eqnarray}
\label{eq:neutrinoeq}
 \dot\rho_\nu+3H(\rho_\nu+p_\nu)=\frac{\beta} {\Mpl}\dot\phi
(\rho_\nu-3p_\nu) \, .
\end{eqnarray}
Let us note that the evolution of the radiation and matter sectors is the
standard one. For the neutrino sector, we have
\begin{eqnarray}
m_{\nu,\rm eff}(\phi)=m_{\nu,0}e^{\beta\phi/\Mpl} \, ,
\label{eq:mnu}
\end{eqnarray}
and hence the neutrino matter behaves as radiation and non-relativistic 
matter during early and late times, respectively.
% where non-minimal coupling builds up.
Using the expression (\ref{eq:mnu}), the effective potential can be re-written as
\begin{eqnarray}
\label{eq:veff}
V_{\rm {eff}}(\phi)= V(\phi) + \left({\rho}_{\nu,0}-3{p}_{\nu,0}\right) e^{\beta\phi/\Mpl}
\,,
\end{eqnarray}
where $\rho_{\nu,0}$ represents the neutrino energy density
with a constant mass, $m_{\nu,0}$, i.e., $\rho_{\nu}=\rho_{\nu,0}e^{\beta\phi/\Mpl}$. Clearly, the coupling massive neutrino matter with field builds up dynamically at late stages as massive neutrinos turn non-relativistic.

As discussed in the Introduction, the unification of  inflation and dark,
energy requires that the field potential should
be shallow, satisfying the slow-roll condition, at the inflationary stage, and steep
after the end of inflation. Since the underlying field potential is
typically of a runaway type that is a characteristic feature of
quintessential inflation, the model belongs to the category of a
non-oscillatory type. Thereby one needs an alternative mechanism
of preheating. E.g. preheating may proceed through the instant
particle production.

In what follows, we shall focus on the generalized exponential potential
\begin{eqnarray}
\label{eq:potential} V(\phi) = V_0 \e^{-\lambda \phi^n/\Mpl^n} \,,
\end{eqnarray}
which  can successfully unify  inflation and  dark energy without
interfering with the thermal history of the universe. If $V(\phi)$ is analytic at the origin, $\phi=0$, that is 
typically expected from a more fundamental microscopic field-theoretical model leading to this effective potential, 
then $n$ is a positive integer. Furthermore, it should be an even positive integer if we want the potential to be 
bounded for all real values of $\phi$. However, all formulas in  Section \ref{sec:inflation} are valid for any $n>2$.
The case $n=2$ requires special consideration, but it cannot produce a good fit to the measured value of the 
slope of the scalar power spectrum $n_s-1$ as follows from Eq. (\ref{eq:spectns2}) in the next Section in the 
limit $n\to 2$. Let us note that the slope of the potential is given by $n \lambda\phi^{n-1}/\Mpl^{n-1}$, and thus 
slow roll is ensured for sufficiently small values of $\phi$ which, however, can much exceed $\Mpl$ for $\lambda\ll 1$. 
This does not lead to any problems in the UV regime since the potential is bounded for even $n$ and can be considered 
as a perturbation for energies $E\gg V_0$, but still much less than the Planck one. As a result, the field acquires the 
approximate shift symmetry at these energies, so all values of $\phi$ are possible.

\begin{figure}[ht]
\centering
\includegraphics[width=0.55 \linewidth]{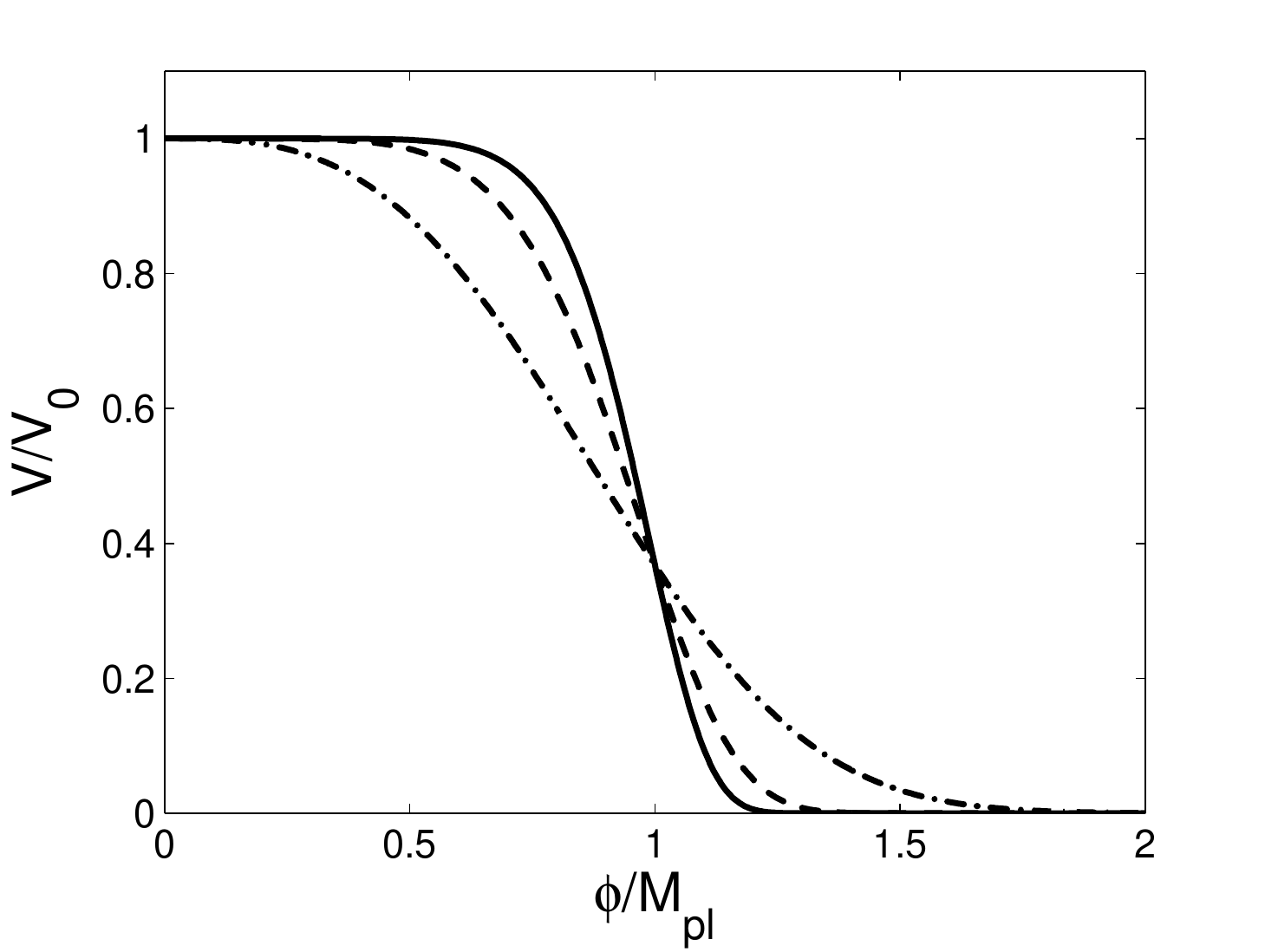}
\caption{{\it{The generalized exponential potential $V(\phi) = V_0 \e^{-\lambda
\phi^n/\Mpl^n}$
with $\lambda=1$, where $n=9$ (solid line), 6 (dashed line) and 3 (dot-dashed line),
respectively.}}
}
\label{fg:5}
\end{figure}

In the steep region away from the origin, the scalar field rolls very fast, and therefore the energy density $\rho_{\phi}$
overshoots the background, but it evolves as a scaling solution at late
times, as $\Gamma \to 1$ for large values of $\phi$. Relativistic
and non-relativistic fluids dominate the universe in this region,
until neutrino-matter becomes non-relativistic that happens at late
times. This then leads to the building of the non-minimal coupling
of $\phi$ with massive neutrino matter, which in turn triggers a
minimum in the effective potential. The slowly rolling scalar field
around the minimum can then mimic the cosmological-constant-like
behavior at late times. There are very few potentials that can
simultaneously pass observational constraints from inflation to
late-time evolution of the universe. In Fig.~\ref{fg:5} this
potential is illustrated with $\lambda=1$, where the solid, dashed
and dot-dashed lines correspond to $n=9$, 6 and 3, respectively. We
observe that when $\lambda \phi^n/\Mpl^n \ll 1$ with a large $n$,
the shallow potential may give rise to  inflation. As $\lambda
\phi^n/\Mpl^n \gtrsim 1$, a scaling solution exists in the steep
region.

In the next Section we will explicitly consider the inflationary stage in this model and confront it with the Planck 2015 results.

\section{Inflation}
\label{sec:inflation}

In the inflationary era, the evolution of the Universe is driven by
the scalar field,  in which case the contributions from $\S_r$,
$\S_m$ and $\S_{\nu}$ should be ignored; they become relevant after
inflation only. Using the standard expressions of slow-roll parameters, namely
\begin{eqnarray}
\label{eq:slow-roll}
\epsilon \equiv \frac{M^2_{\text{Pl}}}{2} \left( \frac{V_{\phi}}{V} \right)^2,
\qquad \eta \equiv M^2_{\text{Pl}} \frac{V_{\phi \phi}}{V}, \qquad \xi \equiv M^ 4_{Pl}
\frac{V_{\phi}V_{\phi \phi \phi}}{V^2}\,,
\end{eqnarray}
we can derive the expressions for the scalar and tensor spectral
indexes ($n_s, n_t$), tensor-to-scalar ratio ($r$) and scalar
spectral index running ($\alpha_s \equiv {\rm d}n_s/{\rm d}\ln k$), as
\begin{eqnarray}
\label{eq:spectns}
&&n_s - 1 = -6\epsilon + 2\eta \,, \\
\label{eq:spectnt}
&&n_t = - 2\epsilon \,, \\
\label{eq:spectr}
&&r = 16\epsilon \,, \\
\label{eq:spectalpha}
&&\alpha_s = 16\epsilon \eta -24\epsilon^2 -2\xi \,.
\end{eqnarray}
The weakly $k$- dependent primordial spectra of scalar and tensor perturbations can be expanded in powers of $\ln(k_/k_s)$ around 
some pivot comoving scale $k_s$:
\begin{eqnarray}
&& \ln\mathcal{P}_s(k) = \ln A_s + (n_s -1) \ln \left(\frac{k}{k_s}\right) + \alpha_s
\left[\ln \left( \frac{k}{k_s}\right) \right]^2 \,, \\
&& \ln\mathcal{P}_t(k) =\ln A_t + n_t \ln \left( \frac{k}{k_s} \right) \,.
\end{eqnarray}
Inserting the potential ~(\ref{eq:potential}) into Eqs.~(\ref{eq:spectns},\ref{eq:spectr}) and using the standard expression for the primordial
scalar spectrum ${\mathcal P}_s$ itself, too, we get
\begin{eqnarray}
\label{eq:spec-phi}
&&{\mathcal P}_s = \frac{V^3}{12\pi^2 \Mpl^6V_{\phi}^2}= \frac{V_0\, \e^{-\lambda\phi^n/\Mpl^n}}
{12\pi^2 n^2 \lambda^2 \Mpl^6\phi^{2n-2}}\,, \\
\label{eq:spectns-phi}
&&n_s - 1 = -n\lambda \left(\frac{\phi}{\Mpl}\right)^{n-2}\left( 2n - 2+ n\lambda \left(\frac{\phi}{\Mpl}\right)^{n}\right)  \,, \\
\label{eq:spectr-phi}
&&r = 8 n^2 \lambda^2 \left(\frac{\phi}{\Mpl}\right)^{2n-2} \, ,
\end{eqnarray}
where, as usually, $\phi(t)$ has to be estimated at the moment $t=t_k$ of the first Hubble radius crossing of a given comoving scale 
$k^{-1}a(t)$ during inflation, i.e. when $k=a(t_k)H(t_k)$. When it follows the observational result $r\lesssim 3(1-n_s)$~\cite{Ade:2015lrj} 
that the range of $\phi$ corresponding to the cosmological scales at which CMB fluctuations are observed lies in the region where the potential
$V(\phi)$ is concave, $V_{\phi\phi}<0$, and where the second term in large round brackets in Eq.~(\ref{eq:spectns-phi}) is significantly less 
than the first one: their ratio 
\begin{equation}
\label{eq:ratio}
\frac{n\lambda}{2n-2}\left(\frac{\phi}{\Mpl}\right)^n = \frac{r}{8(1-n_s)-r}\lesssim 0.6\, . 
\end{equation}
The latter inequality can be made about twice stronger using the more recent combined data from the BICEP2, Keck Array and Planck 
collaborations~\cite{BICEP-Keck}. Note one more useful relation which does not contain the parameter $\lambda$:
\begin{equation}
\label{eq:phi2}
\frac{2\phi^2}{(n-1)^2\Mpl^2}=\frac{r}{\left(1-n_s-\frac{r}{8}\right)^2}\, .
\end{equation}

The usual conditions for the end of inflation are given by either $\epsilon(\phi=\phi_{\rm {end}}) = 1$, or $|\eta |(\phi=\phi_{\rm{end}}) = 1$ 
depending on which of them occurs earlier. So, if $\lambda\ll 1$, we have large-field inflation which ends at
\begin{eqnarray}
\label{eq:phiend}
\phi=\phi_{\rm {end}}= \Mpl \left(\frac{2}{n^2 \lambda^2}\right)^\frac{1}{2n-2} \gg \Mpl\,.
\end{eqnarray}
when both $\epsilon$ and $\eta$ approach unity and 
\begin{equation}
V_{\rm {end}}\ll V_0, ~\ln (V_{\rm {end}}/V_0) \sim -\frac{\phi_{\rm {end}}} {\Mpl} \sim - \lambda^{-\frac{1}{n-1}} \, .
\end{equation}
In the opposite case $\lambda\ll 1$, we get small-field inflation of the hilltop type~\cite{hilltop} which ends when $|\eta|=1,
~\phi=\phi_{\rm {end}}\sim \Mpl \lambda^{-1/(n-2)}\ll \Mpl,~V_{\rm {end}}\approx V_0$. However, only the former case can be used 
for quintessence applications.

The number of the e-foldings counted from the end of inflation can be evaluated as
\begin{eqnarray}
\label{eq:efolding}
\mathcal{N} &=& \int^{t_{\rm{end}}}_{t} H dt^{\prime} = -\Mpl^{-2} \int^{\phi_{\rm{end}}}_{\phi}
\frac{V(\phi^{\prime}) d\phi^{\prime}}{dV(\phi^{\prime})/ d \phi^{\prime}} \\
&=& \frac{1}{n\lambda (n-2)} \left[ \left( \frac{\phi}{\Mpl}\right)^{2-n} - 
\left(\frac{2}{n^2 \lambda^2}\right)^{\frac{2-n}{2n-2}} \right] \,.
\end{eqnarray}
Note the specific and new feature of large-field inflation in our case: the contribution of the upper limit of integration to $\mathcal{N}$ is 
not small compared to unity. Thus, exit from inflation is a rather prolonged one. However, for the values $n=6$ and $n=8$ producing the 
best fit to observational data (see the Table II below) and the corresponding admissible ranges of $\lambda$, this contribution is $\lesssim 4$. 
Thus, it is small compared to the main contribution from the lower limit of integration, $\mathcal{N} = 50 - 60$. Inverting  the relation 
(\ref{eq:efolding}), we acquire
\begin{eqnarray}
\label{eq:phirel}
\frac{\phi}{\Mpl}= \left[ n(n-2)\lambda \mathcal{N} + \left( \frac{2}{n^2
\lambda^2}\right)^{\frac{
2-n}{2n-2}} \right]^{\frac{1}{2-n}} \,.
\end{eqnarray}
%{\color{red}
Using Eqs.~(\ref{eq:slow-roll})-(\ref{eq:spectalpha}) and (\ref{eq:phirel}) with $\lambda \ll 1$, $n>2$ and $\mathcal{N}\gg 1$, one 
can simplify the expressions of $n_s$, $n_t$, $r$, and $\alpha_s$ as functions of  ($\mathcal{N}, n,\lambda$). For
example, the expression of scalar field at commencement of inflation
is reduced to $\phi/\Mpl \simeq \left[n(n-2)\lambda
\mathcal{N}\right]^{-\frac{1}{n-2}}$, while Eqs.~(\ref{eq:spectns}) -
(\ref{eq:spectalpha}) become
\begin{eqnarray}
\label{eq:spectns2}
&&n_s - 1 \simeq -\frac{2(n-1)}{(n-2)\mathcal{N}} - \frac{\left[ n(n-2) \lambda
\mathcal{N} \right]
^{-\frac{2}{n-2}}}{(n-2)^2 \mathcal{N}^2} \lesssim -\frac{2(n-1)}{(n-2)\mathcal{N}} \,, \\
\label{eq:spectnt2}
&&n_t \simeq -\frac{\left[ n(n-2) \lambda \mathcal{N} \right]^{-\frac{2}{n-2}}}{(n-2)^2
\mathcal{N}
^2} < 0 \,, \\
\label{eq:spectr2}
&&r \simeq -8 n_t > 0 \,, \\
\label{eq:spectalpha2}
&&\alpha_s \simeq -\frac{2(n-1)}{(n-2)\mathcal{N}^2} +\frac{6 (n-1) \left[ n(n-2) \lambda
\mathcal{
N} \right]^{-\frac{2}{n-2}}}{(n-2)^3 \mathcal{N}^3} \gtrsim
-\frac{2(n-1)}{(n-2)\mathcal{N}^2} 
\end{eqnarray}
corresponding to  bounds on $n_s$ $r$, $n_t$ and $\alpha_s$, respectively.
%}

\begin{figure}[ht]
\centering
\includegraphics[width=0.49 \linewidth]{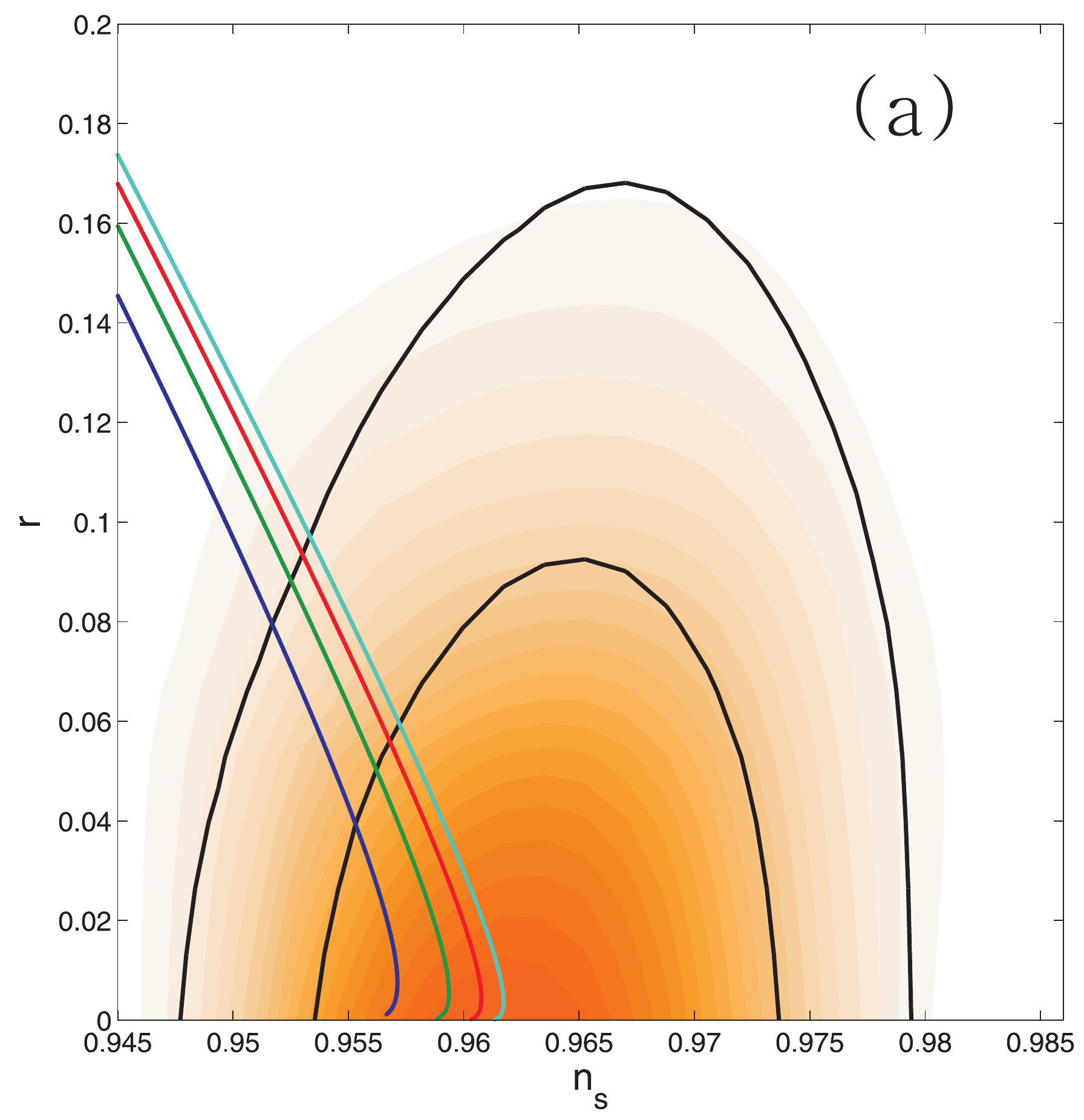}
\includegraphics[width=0.49 \linewidth]{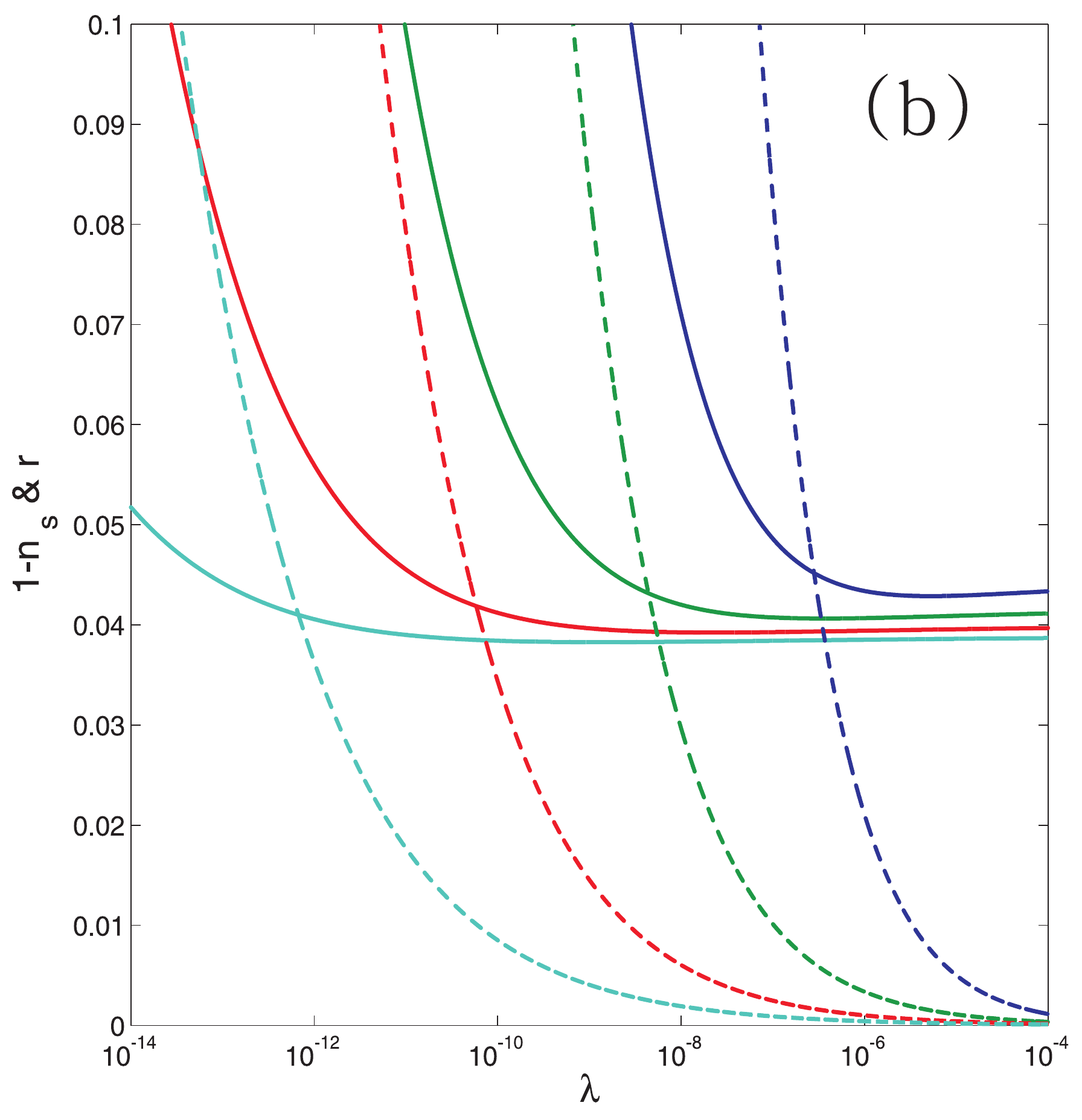}
\caption{{\it{(a) The tensor-to-scalar ratio $r$ as a function of the scalar spectral 
index
$n_s$ in the case of the generalized potential 
$V(\phi) = V_0 \mathrm{exp}(-\lambda\phi^n/\Mpl^n)$,
with $\mathcal{N} = 60$ and $\lambda \leq 10^{-4}$,
where the contours present the $1 \sigma$ and $2 \sigma$ bounds in the
$\Lambda$CDM scenario, respectively.
(b) $1-n_s$ (solid curves) and $r$ (dashed curves) as functions of the
model parameter $\lambda$,
where the blue, green, red and cyan lines correspond to $n=$5, 6, 7 and 8, 
respectively.}}}
\label{fg:1}
\end{figure}

In Fig.~\ref{fg:1} we vary the model parameter $\lambda$ to depict the scalar spectral
index $n_s$
and
tensor-to-scalar ratio $r$ with $V = V_0 \mathrm{exp}(- \lambda \phi^n /\Mpl^n)$ and
$\mathcal{N}=60$.
%Comparing these two plots, we find that the observational constraints
%from the slow-roll parameters yield a lower bound on $\lambda$.
The contour plots in Fig.~\ref{fg:1}a show the Planck 2015 results for the $\Lambda$CDM concordance 
model~\cite{Ade:2015lrj} which yield $n_s \gtrsim 0.952$ and $r \lesssim 0.09$  within $2\sigma$ confidence 
level, and the curves $r(n_s) $ for the generalized exponential potential at hand. In Fig.~\ref{fg:1} the dependence
of $1-n_s$ and $r$ on $\lambda$ is depicted. It is seen that, for a sufficiently large $\lambda$, $n_s -1$ quickly
approaches the $\lambda$-independent value given by the r.h.s. of the last inequality in Eq.~(\ref{eq:spectns2}). 
As a result, a best fit value of $n$  is mainly determined by the measured value of $n_s-1$, while the upper limit
on $r$ produces the lower limit on $\lambda$  that can be seen already from Eq.~(\ref{eq:ratio}). Namely, from  
Eqs.~(\ref{eq:spectns2}) and (\ref{eq:spectr2}) and Fig.~\ref{fg:1}b,
we find that $ \frac{2(n-1)}{(n-2)\mathcal{N}} \lesssim 1-n_s \lesssim 0.048$ and $0
\lesssim r \lesssim 0.09$, which lead to
$\lambda \gtrsim 10^{-7}$, $10^{-9}$, $10^{-11}$ and $10^{-14}$ with $n=5$, 6, 7 and 8,
respectively.
%Clearly, the observational data result in lower bounds on $\lambda$.
%} 
It is clear from the preceding discussion that our model based
upon the generalized potential  (\ref{eq:potential}) provides an accurate description of
inflationary phase. Hence, in the following Section  we shall investigate post-inflationary
evolution.

\section{Post-inflationary evolution} \label{sec:post_inf}

After the end of inflation, the kinetic stage follows when the potential $V(\phi)$ can be neglected. Thus, the 
effective equation of state (EoS) of the field $\phi$ is $p_{\phi}=\rho_{\phi}$. Then
\begin{equation}
\label{eq:kinetic}
a(t)\propto t^{1/3}, ~\phi =\phi_{\rm{end}}+\sqrt{\frac{2}{3}}\Mpl \ln\frac{t}{t_{\rm{end}}},~
\rho_{\phi}=3\Mpl^2H^2=\frac{\Mpl^2}{3t^2} \, .
\end{equation}
To get a transition to the RD stage (the hot Big Bang), usual matter has to be created and heated.
Since the unified potential  (\ref{eq:potential}) is of a runaway type, the standard parametric resonance mechanism is 
not applicable in this case.  However, there exist sufficiently effective alternative mechanisms for this purpose, see 
e.g.~\cite{Felder:1998vq}. In the following discussion we shall study post-inflationary evolution of the field $\phi$ in the 
presence of another background matter with the EoS $p_b=w_b\rho_b,~w_b=const,~0\le w_b<1$ ($w_b=1/3$ at 
the hot RD stage) without invoking a concrete mechanism of its creation and thermalization.

Due to this second kind of matter, the stage~Eq. (\ref{eq:kinetic}) ends at some moment $t=t_b$. After that the Universe 
expands as $a(t)\propto t^q,~q=\frac{2}{3(1+w_b)}$.  Let us first argue on heuristic grounds that there exists an 
approximately scaling late-time solution for $\phi$ in the case of the potential (\ref{eq:potential}). By defining the
dimensionless variables
\begin{eqnarray}
x = \frac{\dot \phi}{\sqrt{6}H\Mpl} \, ,~~ y = \frac{\sqrt{V}}{\sqrt{3}H\Mpl}  \, ,~~
\lambda = -\Mpl \frac{V_{\phi}}{V}\, ,~~ \Gamma= \frac{VV_{\phi
\phi}}{V_{\phi}^{2}},
\end{eqnarray}
we can transform the  equations of motion into the autonomous form
as
\begin{eqnarray}
\frac{{\rm d}x}{{\rm d}N} &=&f(x,y)\, ,
\label{eq:xp}
\frac{{\rm d}y}{{\rm d}N} =
g(x,y) \, ,
\label{eq:Omega_mp} \\
% \frac{{\rm d}w_\nu}{{\rm d}N} &=& \frac{2w_\nu}{ z_{\rm dur}}\(3w_\nu-1\)
%\, ,
% \label{eq:wnup}\\
 \frac{{\rm d}\lam}{{\rm d}N} &=&
-\sqrt{6} \lambda^2 x (\Gamma-1) \, ,
 \label{eq:lamp} \\
\end{eqnarray}
where $N=\ln a$ and $f$ and $g$ are functions of $x$ and $y$ whose explicit forms
are not required for the present discussion. The third equation
becomes redundant in the case of exponential potential. We mention that
the potential (\ref{eq:potential}) is steeper than the standard
exponential function in the post inflationary era (slope $\sim
\phi^{n-1},~n>1$). Moreover, the function $\Gamma$ exhibits a very
interesting feature, namely
\begin{equation}
\Gamma=1-\frac{(n-1)}{n\lambda}\frac{M_p^n}{\phi^n},~~n>1,
\end{equation}
thereby  $\Gamma\to 1$ for large values of the field, and thus the
scaling solution would emerge as an attractor at late times.
\cite{Geng:2015fla}. It is therefore clear that the dynamical system
under consideration mimics scalar field with exponential potential
at late stages.

\subsection{Approximately scaling solution and exit to late-time acceleration}

\label{sec:4_Sol}

In this subsection we shall explicitly demonstrate the existence of the scaling solution.
We shall focus on the scaling behavior in the RD and MD
stages, followed by the dark energy epoch. In the RD and MD epochs the
neutrino masses are negligible and neutrino matter behaves as
radiation, hence $V_{\rm {eff}}(\phi)=V(\phi)$, and thus (\ref{eq:phieq}) results
in the field equation
of the minimally coupled quintessence case, namely
\begin{eqnarray}
\label{eq:phieq2}
\ddot\phi+3H\dot\phi= -\frac{d V}{d \phi}= \frac{n\lambda\phi^{n-1}}{\Mpl^n} \exp\left(-\frac{\lambda \phi^n}{\Mpl^n}\right) \,.
\end{eqnarray}
Using Eqs.~(\ref{eq:FR1}) and (\ref{eq:FR2}) and keeping in mind the
definition
\begin{eqnarray}
\label{eq:eos_rmd}
w_{\phi} \equiv \frac{p_{\phi}}{\rho_{\phi}}=\frac{\frac{1}{2}\dot{\phi}^2 -
V(\phi)}{\frac{1}{2} \dot{\phi}^2 +V(\phi)} \,,
\end{eqnarray}
we obtain the   expression
\begin{eqnarray}
\label{eq:eos_rmd2}
\frac{dV}{d \phi} = \frac{1-w_{\phi}}{1+w_{\phi}} \ddot{\phi} \,,
\end{eqnarray}
where $w_{\phi}$ has been assumed to be constant in view of the
scaling solution we are interested in.
For the background dominated fluid with equation of state $w_b$, the
Hubble parameter has the form  $H=2/3(1+w_b) \cdot t^{-1}$.
Substituting Eq.~(\ref{eq:eos_rmd2}) into (\ref{eq:phieq2}), one
readily obtains
\begin{eqnarray}
\label{eq:phieq3}
\ddot{\phi}+\frac{3H (1+w_{\phi})}{2}\dot{\phi}=0 \,,
\end{eqnarray}
{ such that for $w_\phi=const$, $\dot{\phi} \sim t^{p}$ is a solution of Eq.~(\ref{eq:phieq3}),
where $p$ is a real number. However, this solution does
not satisfy the field equation, since the time exponents of different
terms in the equation do not match with each other which is not surprising as $w_\phi$ is not constant in general}. Instead, the asymptotic late-time solution 
of Eqs.~(\ref{eq:phieq2}) for the potential
(\ref{eq:potential}) 
has the form of the following series:
\begin{eqnarray}
\label{eq:phisol}
\lambda\(\frac{\phi}{\Mpl} \)^n = C_1 \ln (H_1 t) + C_2 \ln \ln (H_1 t) + ...\,,
\end{eqnarray}
where $C_1$, $C_2$ and $H_1$ are constant parameters to be determined. Here, we assume $\lambda \ll 1$, so that 
the scalar field undergoes the scaling regime with $\phi \gg \Mpl$ and $\ln(H_1t)\gg 1$ (note that $H_1$ is {\it not} of the 
order of $t_b$ as is seen from the expression Eq.~(\ref{eq:coef}) below).

After substituting the {\it Ansatz}~(\ref{eq:phisol}) into Eq.~(\ref{eq:phieq2}), the latter equation becomes
\begin{equation}
\label{eq:phisol2}
\frac{(3q-1)C_1^{\frac{1}{n}} \lambda^{-\frac{1}{n}}\Mpl \, (\ln(H_1t))^{\frac{1-n}{n}}}{nt^2} = \frac{nC_1^{\frac{n-1}{n}}
\lambda^{\frac{1}{n}}V_0 \, (\ln(H_1t))^{\frac{n-1}{n}-C_2}}{\Mpl (H_1t)^{C_1}}
\end{equation}
in the leading approximation. Comparing the powers and coefficients in the LHS and RHS of Eq.~(\ref{eq:phisol2}), we 
find that 
\begin{equation}
\label{eq:coef}
C_1=2,~C_2=2 - \frac{2}{n},~ H_1^2= \frac{2^{\frac {n-2}{n}}n^2\lambda^{\frac{2}{n}}V_0}{(3q-1)\Mpl^2}\, .
\end{equation}
%where the first term in before the parenthesis reminds us the fractional density parameter
%for the standard exponential potential with slope $\lambda$. 
%It is interesting to note that expression (\ref{eq:rhophiratio}) has the same 
%meaning in the generalized case (\ref{eq:potential}) with $n \gg 1$. Indeed,
%\begin{eqnarray}
%\Omega_{\phi}\simeq \frac{4}{n^2} \frac{1}{(\lambda\phi^n/\Mpl^n)^2} \,.
%\end{eqnarray}
%where $n \lambda \phi^n/\Mpl^n$ is the slope of the generalized exponential potential (\ref{eq:potential}) with $n \gg 1$.
%where $C_1$ and $C_2$ are constant parameters with $\lambda C_1=2$,
%leading to $V(\phi) \propto t^{-2}$. As a result, the powers of $t$
%in  all terms in Eq.(\ref{eq:phieq2}) are consistent with each
%other, and
In addition, it is easy to check that
\begin{eqnarray}
\frac{\ddot{\phi}}{\dot{\phi}} \approx  -\frac{1}{t}- \frac{n-1}{nt\ln(H_1t)}
\xrightarrow{\quad \lambda \phi^n/\Mpl^n \gg 1 \quad} -\frac{1}{t} \,,
\end{eqnarray}
which clearly satisfies Eq.~(\ref{eq:phieq3}) provided that we make the following identification:
\begin{eqnarray}
\label{eq:phiw}
w_{\phi}= w_b \,.
\end{eqnarray} 
In fact, the equation of state parameter for the field in the asymptotic regime (\ref{eq:phisol}) has the following form,
\begin{eqnarray}
w_{\phi} =  w_b +{\mathcal O}\left(\frac{\ln\ln(H_1t)}{\ln(H_1t)}\right) \, ,
\end{eqnarray}
which shows that the system gradually settles to the approximately scaling regime.
As a result, the relative field energy density ratio at this regime can be solved as
\begin{equation}
\label{eq:rhophiratio}
\Omega_{\phi}= \frac{\rho_{\phi}}{3\Mpl^2H^2}\approx \frac{V(\phi) t^2}{q(3q-1)\Mpl^2}=\left(2^{\frac{n-2}{n}}qn^2
\lambda^{\frac{2}{n}}\, (\ln(H_1t))^{2-\frac{2}{n}}\right)^{-1} = \frac{2}{qn^2\lambda^2}\left(\frac{\Mpl}{\phi}\right)^{2n-2} \, .
\end{equation}
For $n=1$ (the standard exponential potential), this expression reduces to the well known result $\Omega_{\phi}= 
\frac{2}{q\lambda^2}=const$.

It follows from Eq.~(\ref{eq:rhophiratio}) that the approximately scaling regime begins at the moment $t=t_s$ when $\ln(H_1t) \sim  \lambda^{-\frac{1}{n-1}}\gg 1$ and $\phi\sim \Mpl \lambda^{-\frac{1}{n-1}}\sim \phi_{\rm {end}}$, where $\phi_{\rm {end}}$ 
was defined in Eq.~(\ref{eq:phiend}). However, in fact $t_s\gg t_b$, so it cannot begin immediately after the moment when 
$\rho_{\phi}=\rho_b$. The reason for this is that the regime  Eq.~(\ref{eq:rhophiratio}) requires$\frac{\dot\phi^2}{2V(\phi)}=(1+w_b)/(1-w_b)=1/(3q-1)\sim 1$, while $V(\phi)\ll \dot\phi^2$ during the kinetic stage 
Eq.~(\ref{eq:kinetic}), and at $t=t_b$ in particular. Thus, another, a rather short intermediate stage should occur between the kinetic and 
scaling regimes until $\rho_b\propto t^{-2}$ falls down to the value of $V(\phi)$. At this stage
\begin{equation}
\label{eq:intermediate}
a(t) \propto t^q,~q > \frac{1}{3},~\dot\phi \propto a^{-3}\propto t^{-3q},~\phi \approx const,~V(\phi)\approx V_1= const \, .
\end{equation}
Thus, 
\begin{equation}
t_s\sim \frac{\Mpl}{\sqrt{V_1}} \sim t_b \sqrt{\frac{\rho_{\phi}(t_b)}{V_1}}\gg t_b \, . 
\end{equation}

\begin{figure}[ht]
\centering
\includegraphics[width=0.49 \linewidth]{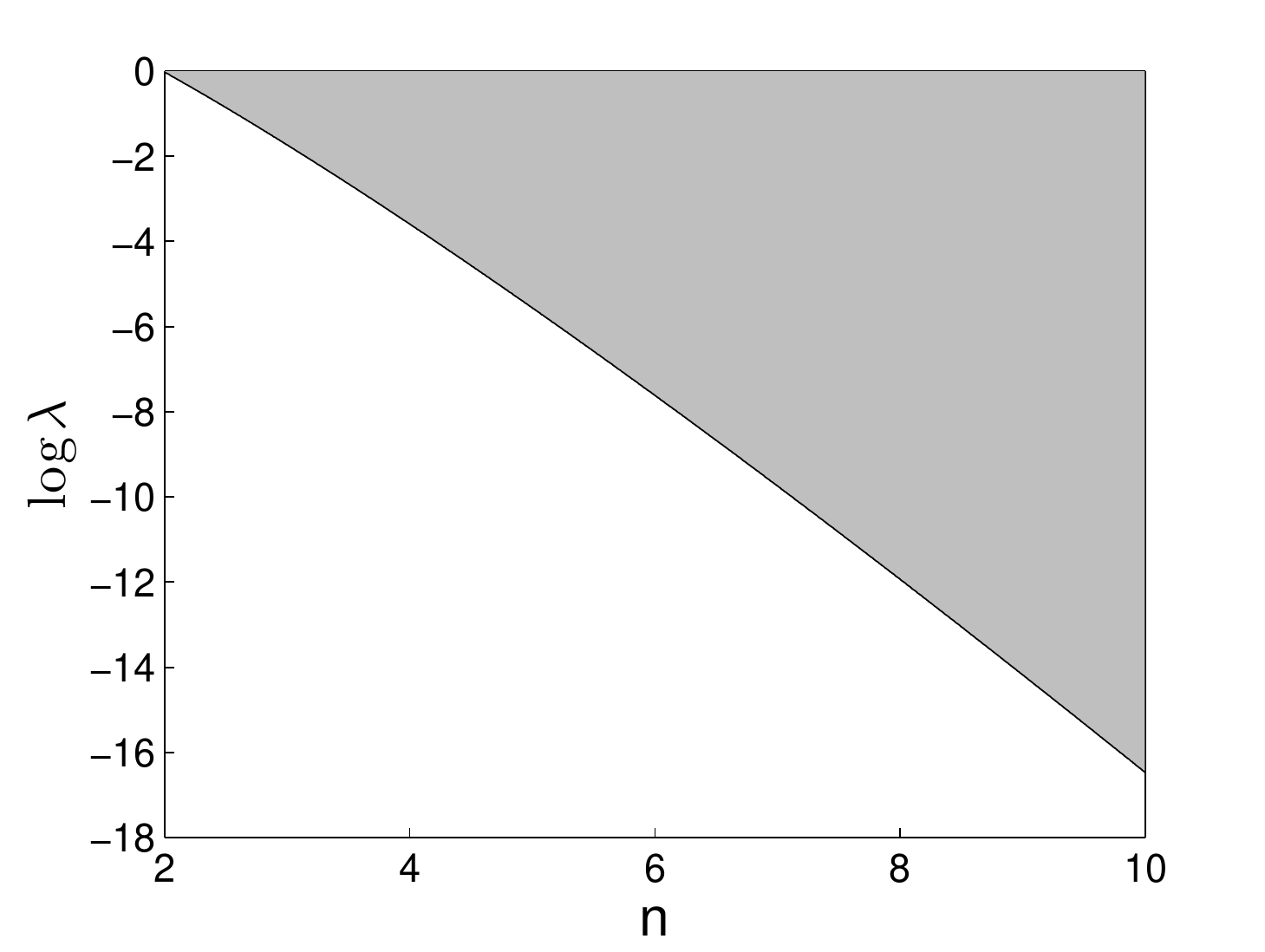}
\caption{{\it{The gray region shows the allowed parameter space from the recombination constraint.}}}
\label{fg:6}
\end{figure}

The fact that scaling at the stage (\ref{eq:phisol},\ref{eq:rhophiratio}) is only an approximate one in our model is crucial since it leads
to $\Omega_{\phi}\ll 1$ at the RD and MD stages (apart from a short period around $t=t_s$ at the beginning of scaling). This provides
a possibility to satisfy observational upper limits on the amount of early dark energy (quintessence in our case) at the Big Bang 
Nucleosynthesis (BBN) period in the early Universe ($q=1/2$) and at the recombination moment at the MD stage (the redshift 
$z\approx 1100$ when $q\approx 2/3$).
%If the scalar field is assumed to leave the frozen stage and evolve to the scaling regime
%with $\lambda \phi^n = \mathcal{K}(t)$ before the end of the Big Bang Nucleosynthesis
From BBN, we have the constraint $\Omega_{\phi}<0.045$  \cite{Bean:2001wt}. The most recent result on the primordial $^4$He 
abundance~\cite{Izotov14} 
 leads to  the effective neutrino number $\Delta N_{eff}$ to be $0.53 \pm 0.50$ at $99\%$ confidence level, corresponding to $\Omega_{\phi} < 0.11$ with $N_{eff}=3.046$.
%, where we have used $\Omega_{\phi} = \frac{3}{4} \frac{7 \Delta N_{eff}/4}{10.75+7 \Delta N_{eff}/4}$~\cite{Ferreira:1997au}.} 
The upper limit from recombination is an order of magnitude stronger: $\Omega_{\phi}<0.0036$ in Ref.~\cite{Ade1590}. 
Consequently,  from Eq.~(\ref{eq:rhophiratio}) we find that
\begin{eqnarray}
\label{eq:rhophiratioBBN}
\lambda n (\phi/\Mpl)^{n-1}_{BBN} > 6.03 \quad \mathrm{and} \quad \lambda n (\phi/\Mpl)^{n-1}_{rec} > 28.9 \,.
%\left. \Omega_{\phi} \right|_{BBN}\simeq \left. \frac{4}{n^2} \frac{1}{(\lambda
%\phi^n/\Mpl^n)^2} \right|_{BBN} \simeq \frac{4}{n^2 \mathcal{K}_{BBN}^2}<0.045\,.
\end{eqnarray}
%Consequently, the result of {\color{red}$n > ... $} indicates that the
%generalized exponential potential passes the BBN and recombination constraints with large enough $\mathcal{K}_{BBN}$ and $n$.
Subsequently,
by combining with the numerical values of
$\lambda (\phi/\Mpl)^n_{BBN} \simeq 170 $ and $\lambda (\phi/\Mpl)^n_{rec} \simeq 220$ with $V_0/H_0^2 \Mpl^2 = 10^{100}$, 
the second condition in Eq.~(\ref{eq:rhophiratioBBN}) is reduced to
\begin{eqnarray}
\label{eq:lamnconstr}
n \lambda^{\frac{1}{n}} > 28.9 \times (220)^{\frac{1}{n}-1} \,.
\end{eqnarray}
Clearly, Eq.~(\ref{eq:lamnconstr}) is always valid when $n \gg 1$, while the allowed parameter space for  small numerical values of $n$ is shown in Fig.~\ref{fg:6}.
%Note that the numerical values with $V_0/H_0^2 \Mpl^2 = 10^{100}$ show that
%$\lambda (\phi/\Mpl)^n_{BBN} \simeq 170 $ and $\lambda (\phi/\Mpl)^n_{BBN} \simeq 220$, which are independent on $\lambda$ and $n$.

Now let us consider transition from the MD stage to the stage of dark energy dominance which occurs at the present epoch.
In the absence of non-minimal coupling of neutrinos to quintessence ($\beta =0$ in Eq. (\ref{eq:coupling})), the approximately 
scaling MD stage would  run forever. A non-zero $\beta$ changes the effective potential by adding a $\phi$-dependent term to it 
according to Eq.(\ref{eq:veff}). As follows from the relation (\ref{eq:phisol}), the scalar field
grows in time during the  RD and MD epochs, and the growth of $\phi$ results
in the increase of neutrino masses. With a large enough $\phi$,
neutrino-matter turns into non-relativistic, and the neutrino
contribution in the effective potential (\ref{eq:veff}) is no longer
negligible. Hence, the coupling to the field builds up, giving rise to the minimum
in the effective potential; the original potential is of a steep run-away type. As the 
scalar approaches the minimum of the effective
potential, it oscillates around it and finally settles down in the
minimum at $\phi_m$, and the universe in the scenario under consideration enters the 
dark-energy dominated
stage. In this case, the
numerical value of the scalar at   present time, namely $\phi_0$, is approximately equal to
$\phi_m$, i.e.
\begin{eqnarray}
\label{eq:phimin}
\phi_0 \simeq \phi_m = \Mpl\left[ \frac{\ln \left( V_0/ \rho_{\phi,
0}\right)}{\lambda}\right]^{\frac{1}{n}} \,,
\end{eqnarray}
derived by assuming $\rho_{\phi, 0} = \dot{\phi}^2 /2+V(\phi) \simeq V(\phi)$, where
$\rho_{\phi, 0}
$ is the energy density of the scalar at  present time.
Combining (\ref{eq:phimin}) with
\begin{eqnarray}
\frac{ d V_{\rm {eff}}}{ d \phi} \rvert_{\phi=\phi_m} =0\,,
\end{eqnarray}
we obtain
\begin{eqnarray}
\label{eq:beta}
\beta =  \frac{\lambda n \phi_0^{n-1} \rho_{\phi, 0}}{\Mpl^{n-1}\rho_{\nu, 0}} \,.
\end{eqnarray}

\begin{figure}[ht]
\centering
\includegraphics[width=0.49 \linewidth]{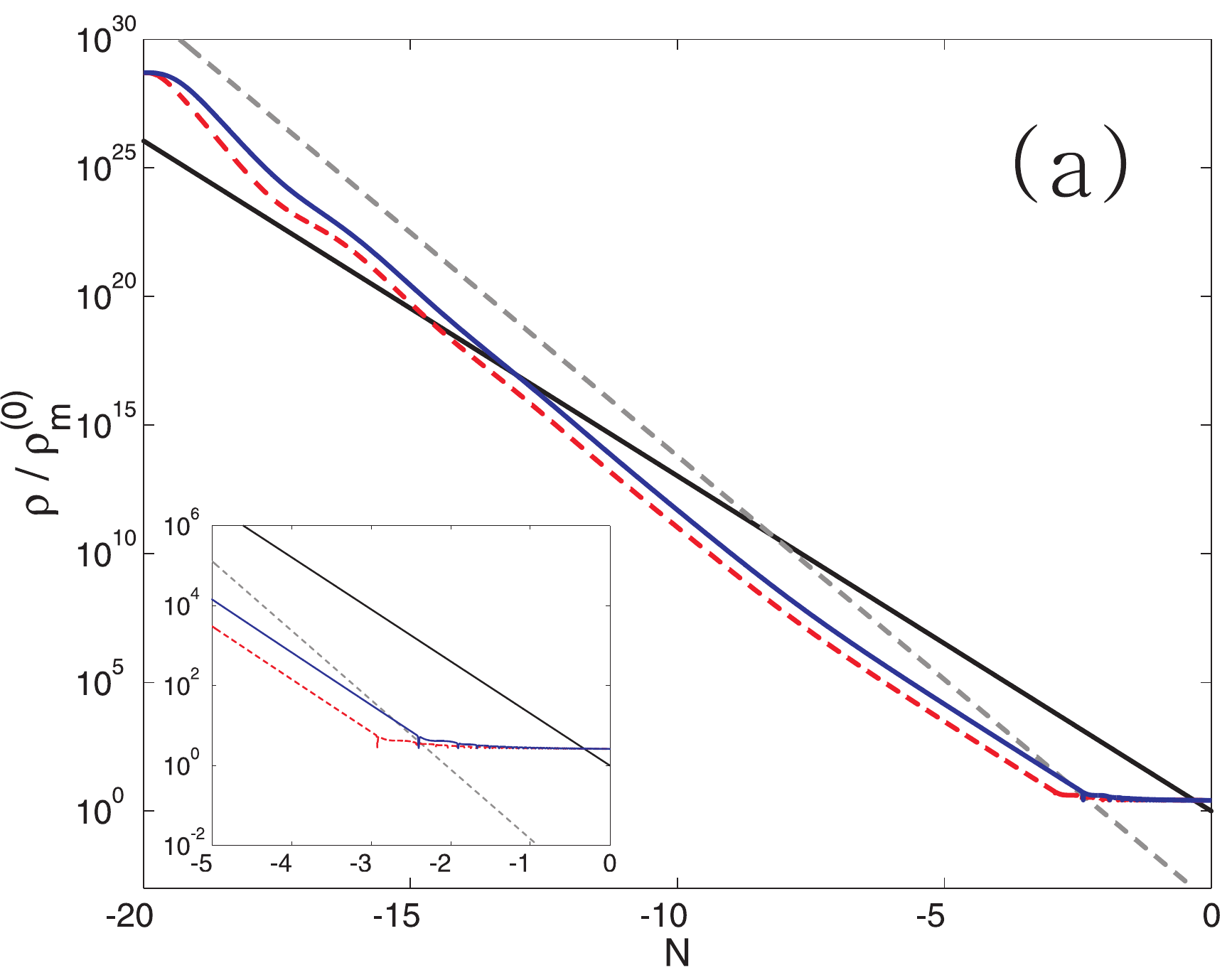}
\includegraphics[width=0.49 \linewidth]{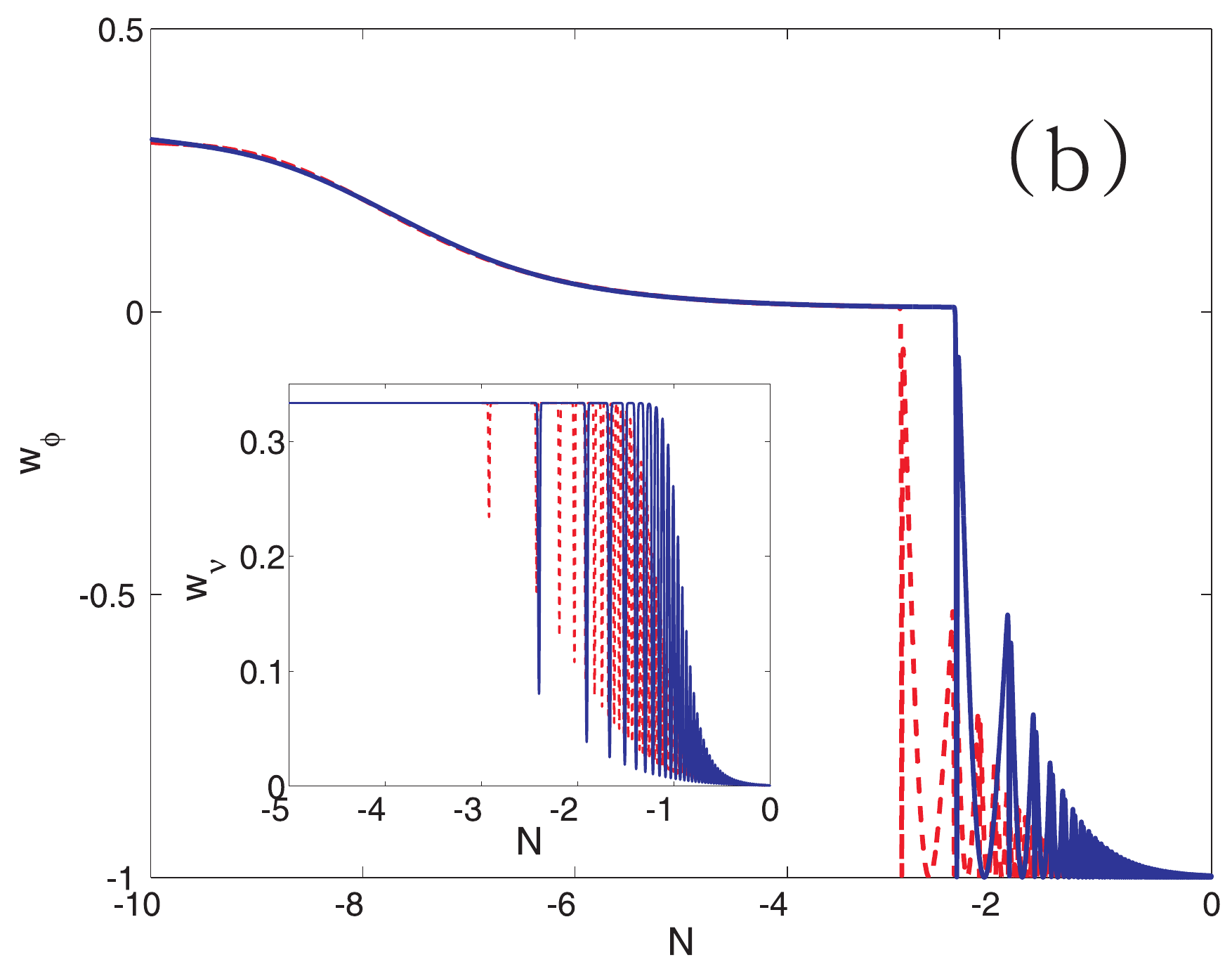}
\caption{{\it{
(a) Energy densities of $\rho_r$ (gray-dashed), $\rho_m$ (black-solid) and
$\rho_{\phi}$ with $\lambda$= $10^{-8}$ (blue-solid) and $10^{-6}$ (red-dashed), 
normalized
by the
matter energy density $\rho_{m}^{(0)} \equiv \rho_m \lvert_{z=0}$, as functions of $N
\equiv \ln a$,
with the potential $V(\phi) = V_0 \mathrm{exp}\left(-\lambda (\phi^n/\Mpl^n) \right)$
%.
%$\rho_r$ is the gray-dashed line, $\rho_m$ is the black-solid line, and $\rho_{\phi}$ is
plotted
for $\lambda$= $10^{-8}$ (blue-solid) and $10^{-6}$ (red-dashed) and $20$ (red-dotted),
with
$\Omega_c h^2 =0.118$.
(b) The equation-of-state parameters $w_{\phi}$ and $w_{\nu}$  as functions of $N$,
where we have used $\Sigma m_{\nu} = 0.45$~eV, $\Omega_m h^2 =0.118$ and $\rho_r^{(0)}/
\rho_m^{(0)}
=2.6\times 10^{-4}$ as   boundary conditions.
}}}
\label{fg:2}
\end{figure}

In Fig.~\ref{fg:2}, we numerically solve the evolution equations to demonstrate the evolution history.
Since the cosmic evolution after the post-inflationary epoch is insensitive to $V_0$,  in our numerical
calculations we choose $V_0/\rho_{m, 0} = 10^{110}$, corresponding to the inflationary scale.
In Fig.~\ref{fg:2}a, the energy density $\rho_{\phi}$ exhibits a tracker scaling behavior in RD and
MD epochs, and the dark energy dominated stage occurs after the
neutrinos become massive. Fig.~\ref{fg:2}b shows the evolutions of
$w_{\phi}$ and $w_{\nu}$. The EoS of the
$\phi$-field behaves as radiation (matter) in the RD (MD) epoch,
which is consistent with the result of relation (\ref{eq:phiw}). After
the neutrino mass becomes important, $\phi$ reaches its minimum
$\phi_m$ and oscillates around it at $z \lesssim 3$. This plot also
demonstrates that the neutrino mass is negligible ($w_{\nu} = 1/3$)
in early times ($z \gtrsim 10$); oscillates when $\phi$ evolves
around $\phi_m$ in the intermediate region ($ 0 \lesssim z \lesssim
3$), and finally acquires its present value when $w_{\nu}=0$ at
late-times, in which  dark energy is dominant.

Finally, we would like to mention that the scalar $\phi$ within a large-scale neutrino 
lump  deviates significantly from that at the background level, leading the neutrino masses 
to be negligible inside the neutrino lump, and the perturbation becomes 
non-linear~ \cite{Mota:2008nj, Ayaita:2014una}.
In this work we avoid such a non-linear region and  we assume that the neutrino mass 
$m_{\nu}$ and the scalar $\phi$ are both homogeneous, and thus the perturbation of 
neutrinos in the non-minimally-coupled 
scenario behaves in the same way as that in $\Lambda$CDM cosmology, namely
\begin{eqnarray}
\dot{\delta}_{\nu} = 3 H \left( w_{\nu} - \frac{\delta p_{\nu}}{\delta \rho_{\nu}} 
\right) \delta_{\nu} - (1+w_{\nu})\left( \theta - \frac{\dot{h}}{2} \right) \,.
\end{eqnarray}

In summary, from the above discussions we deduce that the non-minimally coupled 
neutrino-matter scenario, with the
generalized exponential potential, perfectly describes  the universe evolution history.
In the next subsection, we will confront this scenario with  cosmological observational data.

\subsection{Observational constraints} \label{subsec:constrant}

We use the CosmoMC program  \cite{Lewis:1999bs, Lewis:2002ah} in order to extract 
observational
constraints on the quintessential inflation scenario.
In our analysis, we include the data of the cosmic microwave background (CMB) from
Planck~ \cite{Adam:2015wua, Aghanim:2015xee}, baryon acoustic oscillation (BAO)  from 
Baryon
Oscillation
Spectroscopic Survey (BOSS)~ \cite{Anderson:2012sa,Anderson:2013zyy}, and Type-Ia
supernova (SNIa)
from Supernova Legacy Survey (SNLS)~ \cite{Astier:2005qq}. The details of the
fitting procedure can be found in Refs.~ \cite{Lewis:1999bs, Lewis:2002ah}.
The prior for the parameters is listed in Table.~\ref{table:1}.
\begin{table}[ht] 
\begin{center}
\begin{tabular}{|c||c|} \hline
Parameter & Prior\\
\hline
Neutrino mass sum& $0.1 \leq \Sigma m_{\nu} \leq 4.0$~eV
\\ \hline
Model parameter $\lambda$ ($n=6$) & $-10 \leq \log \lambda \leq -5$
\\ \hline
Model parameter $\lambda$ ($n=8$) & $-14 \leq \log \lambda \leq -7$
\\ \hline
Baryon density & $0.5 \leq 100\Omega_bh^2 \leq 10$
\\ \hline
CDM density & $10^{-3} \leq \Omega_ch^2 \leq 0.99$
\\ \hline
\end{tabular}
%\vskip 0.2in
\label{table:1}
\caption{%\color{red}
Priors for cosmological parameters with $V(\phi) =V_0 e^{-\lambda (\phi / \Mpl)^n}$.  }
\end{center}
\end{table}

\begin{figure}[ht]
\centering
\includegraphics[width=0.99 \linewidth]{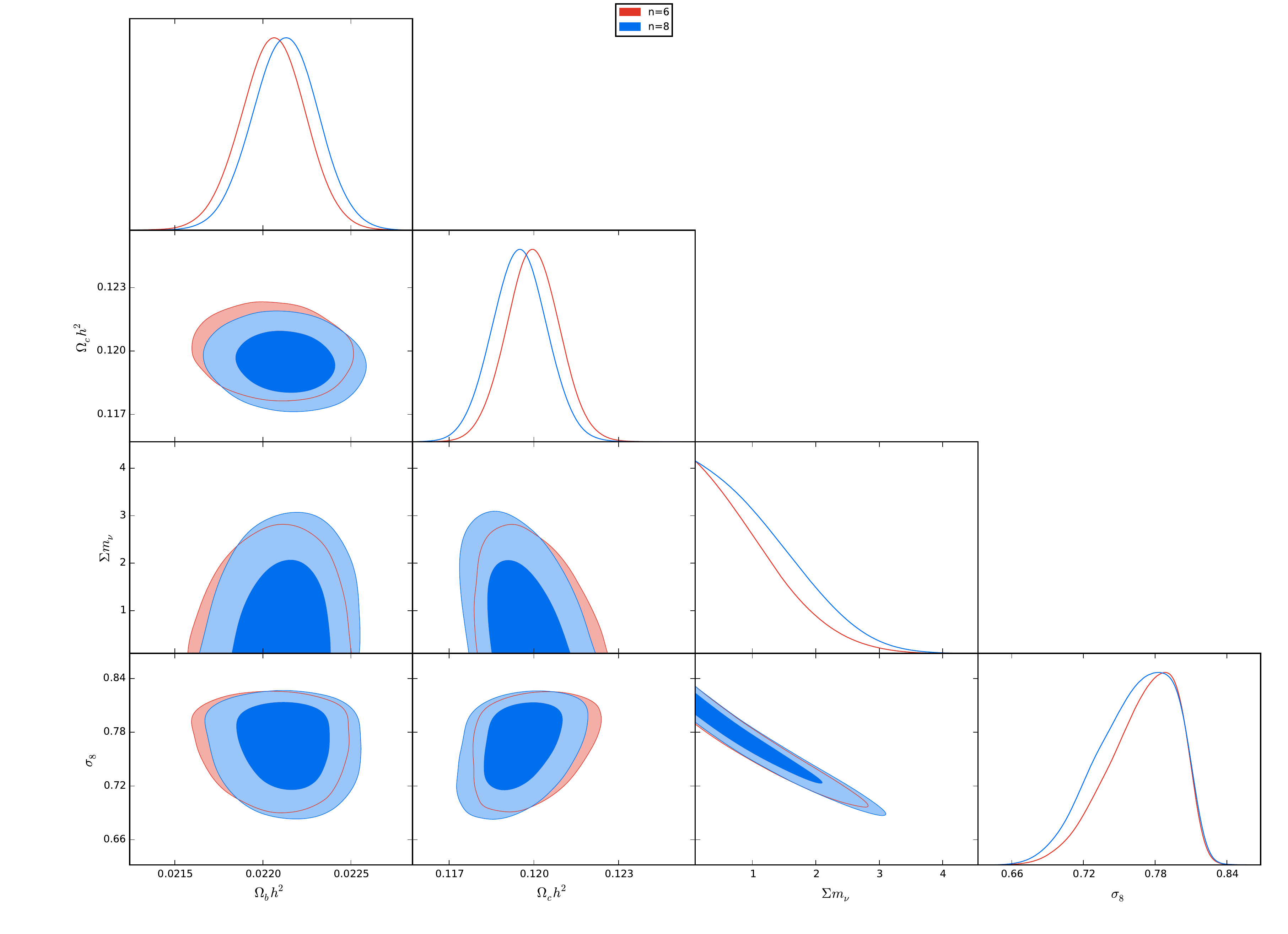}
\caption{{\it{One and two-dimensional distributions of $\Omega_b h^2$, $\Omega_c h^2$, 
$\sum 
m_{\nu}$ 
and $\sigma_8$, for $n=6$ (orange) and $n=8$ (blue), where the
contour lines represent the 68\% and 95\% confidence levels, respectively.}}}
\label{fg:3}
\end{figure}

To prevent non-analytical behaviour at $\phi=0$ and to have a maximum of $V(\phi)$ at this point, we focus on 
even values of $n$. It appears that the values $n=6$ and $n=8$ produce the best fits to observational data.  In Fig.~\ref{fg:3}, 
we depict the $2D$ likelihood contours for  $\Omega_b h^2$, $\Omega_c h^2$, $\sum m_{\nu}$ and $\sigma_8$ with 
$n=6$ (orange) and $n=8$ (blue), where the
contour lines represent 68\% and 95\% confidence levels, respectively.
The scenario at hand is in agreement with observations, in which the $\chi^2$ values 
for $n=6$ and $8$ are equal and even less than that of the $\Lambda$CDM model \footnote{Note for completeness that if we omit
the assumption of $n$ being even and, purely phenomenologically, permit it to be any real number, we get $ n =6.74^{+1.08}_{-0.59}$
($68\%$ confidence limits).}.
The cosmological quantities, such as the present dark-matter and dark-energy densities,
have similar ranges as those in  
$\Lambda$CDM cosmology~ \cite{Adam:2015wua, Aghanim:2015xee}.
However, the neutrino mass sum in the model, in both $n=6$ and $8$ cases, is enhanced to 
$1$~eV, and 
the allowed
window is also
significantly relaxed,
such that $\Sigma m_{\nu} \lesssim 2.5$~eV.
\begin{figure}[!]
\centering
\includegraphics[width=0.39 \linewidth]{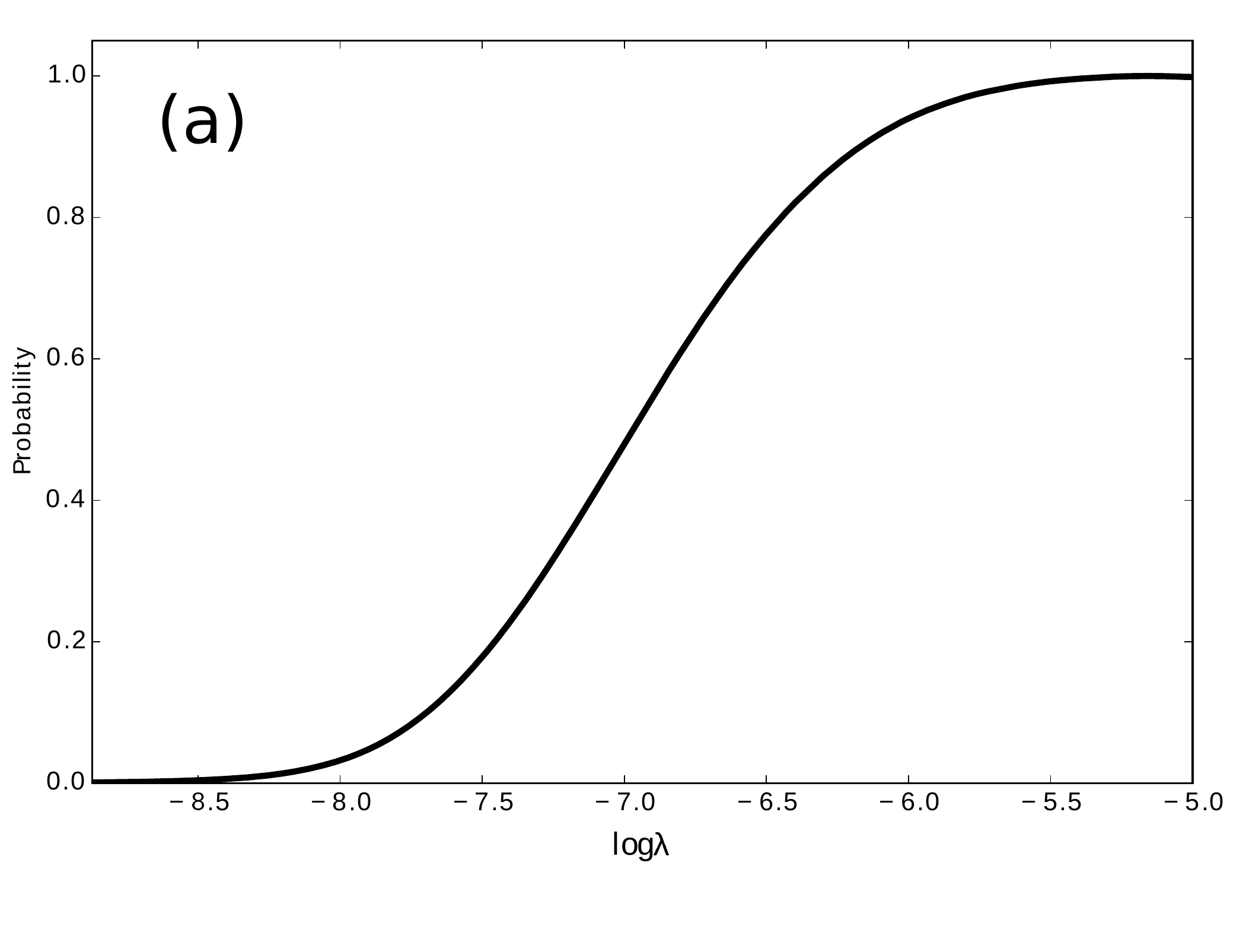}
\includegraphics[width=0.39 \linewidth]{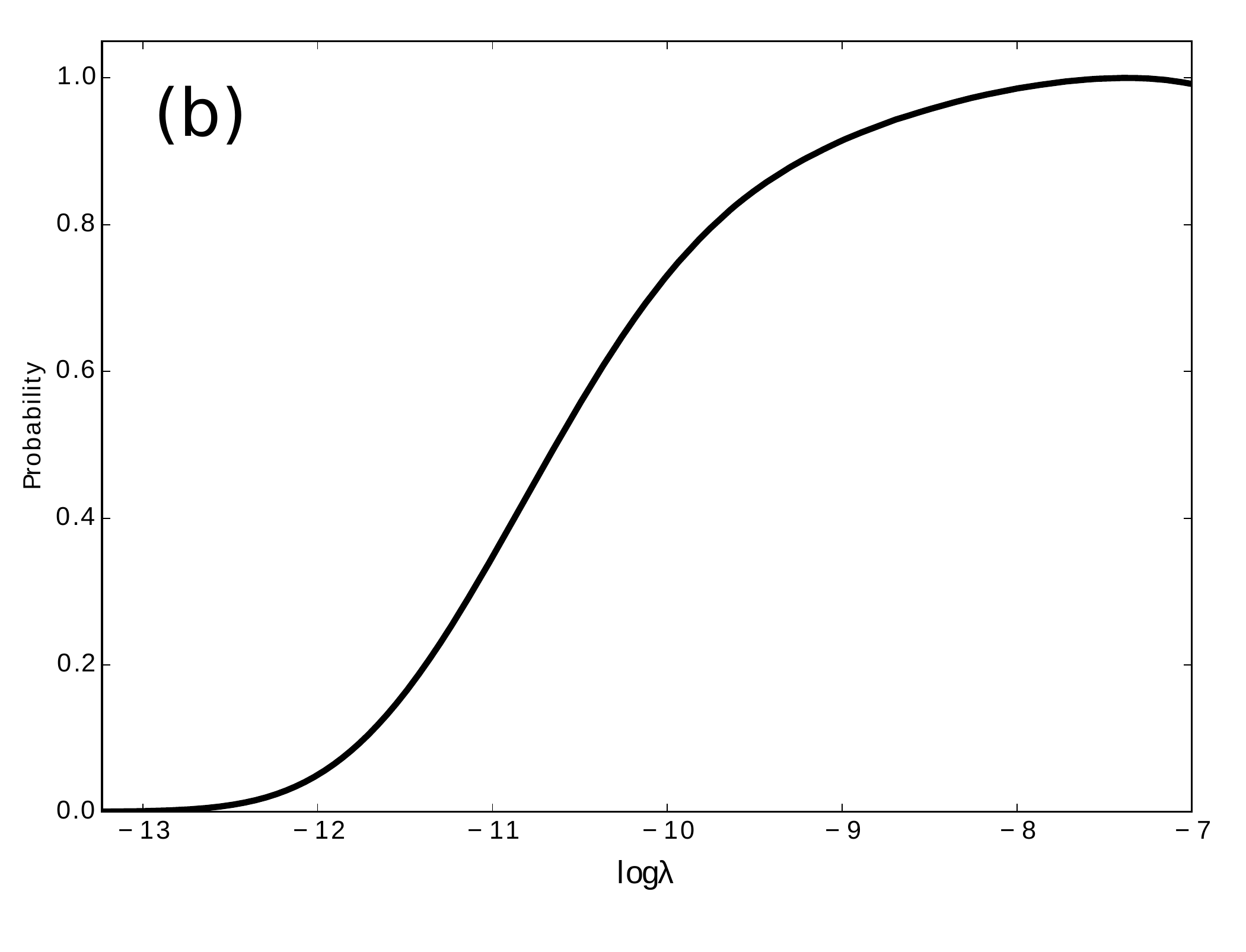}
\caption{{\it{Marginalized probabilities for the potential parameter $\lambda$ with (a) 
$n=6$ 
and (b) 
$n=8$, respectively.}}}
\label{fg:4}
\end{figure}

As discussed in Sec.~\ref{sec:inflation}, the model parameter $\lambda$ is
bounded from below. However, this bound depends on the other parameter $n$ and it is smaller than $10^{-10}$ 
when $n
\gtrsim
7$.
%To avoid the numerical problem,  instead of searching the lower bound, we concentrate on
%the preferred parameter space,
%constrained from the observational data.
In Fig.~\ref{fg:4}, we present the $1D$ marginalized probability plot for $\lambda$ with 
$n=6$ and 8.
As we can see, the cosmological data prefer a larger value of $\lambda$, 
corresponding to 
$n_s \rightarrow 0.96$ and $r \rightarrow 0$ in both cases.
Finally, it is worth to mention that the coupling $\beta$ is tuned to control the values
of $\phi_
0$ and $\Omega_{\phi}$ in the numerical analysis (for instance, from 
Eqs.~(\ref{eq:phimin}) and (\ref{eq:beta}), we have $ (\phi_0, \beta)
\simeq (54,
1800)$ with $(\lambda, n)=(10^{-8}, 6)$).

\begin{table}[ht]
\begin{center}
\begin{tabular}{|c||c|c|c|} \hline
Parameter &  Quintessential Inflation (n=6) &  Quintessential Inflation (n=8) & 
$\Lambda$CDM
\\ \hline
Spectral index & $ n_s = 0.960 \pm 0.001$ & $ n_s = 0.961 \pm 0.001$ &
$ n_s = 0.970 \pm 0.009 $
\\ \hline
Tensor-to-scalar ratio & $r<1.72 \times 10^{-2}$ & $r<2.32 \times 10^{-2}$ & $r<0.125$
\\ \hline
Baryon density  & $ 100 \Omega_bh^2 =   2.21 ^{+0.03}_{- 0.04}$ & $ 100 \Omega_bh^2 =   
2.21 ^{+0.
04}_{- 0.03}$ & $ 100 \Omega_bh^2 =
2.23\pm0.04$
\\ \hline
CDM density  & $ \Omega_ch^2 =  0.120 \pm 0.002$ & $ \Omega_ch^2 =  0.119 ^{+0.002}_{- 
0.001}$ & $ \Omega_ch^2 = 0.118 ^{+0.002}_{- 0.003}$
\\ \hline
Neutrino mass  & $\Sigma m_{\nu} =0.96^{+1.29}_{-0.86}$~eV & $\Sigma m_{\nu} 
=1.13^{+1.41}_{-1.03}
$~eV & $\Sigma m_{\nu} <
0.24$~eV
\\ \hline
Model parameter $\lambda$ & $ \log \lambda > -7.29 $ & $ \log \lambda > -11.7$ & $-$
\\ \hline
\vspace{0.1mm}
$\sigma_8$  & $\sigma_8= 0.770^{+0.043}_{-0.060}$ & $\sigma_8= 0.763^{+0.057}_{-0.062}$ & 
$\sigma_
8=0.805 \pm 0.027$ 
\\ \hline
$\Delta \chi^2 \equiv \chi^2_{q} - \chi^2_{\Lambda CDM}$  & $\Delta \chi^2 = +0.2$ & 
$\Delta \chi^2 
= -2.8$ & $-$
\\ \hline
\end{tabular}
%\vskip 0.2in
\label{table:2}
\caption{%\color{red} 
List of allowed regions with $95\%$ C.L., in the case of   $V(\phi) =V_0 e^{-\lambda 
(\phi / 
\Mpl)^n}$.  }
\end{center}
\end{table}

Our detailed results for the non-minimally-coupled scenario are summarized in
Table.~\ref{table:2}.

\section{Conclusions} \label{sec:conclusion}

In this work we have investigated a model of   quintessential inflation based
upon the generalized exponential potential $V=V_0\exp\left(-\lambda
\phi^n/\Mpl^n \right),~n>1$. This simple function has remarkable
properties, namely its slope behaves as $\phi^{n-1}$, and thus it can
facilitate slow roll for $\lambda \phi^{n-1}/\Mpl^{n-1} \ll 1$ (if $\lambda \ll 1$). For large values of the field, 
$\lambda \phi^{n-1}/\Mpl^{n-1}\gtrsim 1$, the parameter $\Gamma=V_{\phi\phi}V/V_{\phi}^2\to 1$, and as a 
consequence  the system effectively mimics scaling behavior at late stages. There is a
comfortable parameter space in $(n,\lambda)$ that allows us to
satisfy the observational constraints related to inflation. In
Fig.~\ref{fg:1} we have displayed the $2\sigma$  contours based
upon Planck data and the predictions of the scenario at hand, which show that the
model performs as desired for $n\ge 5$ and $\lambda \le 10^{-4}$. 
%It should be noted that the model under consideration exhibits the properties of small-field {\it hilltop inflation} \cite{hiltop,Ade:2015lrj}. 
After the end of inflation, the field enters
into the steep regime of the potential. It was indicated  earlier that the generalized potential gives
rise to an approximately scaling behavior despite being steeper than the standard
exponential one. We have analytically shown here the existence of
approximately scaling solutions in the followed-up radiation and matter dominated
epochs. At for the late time evolution, we need a mechanism of exit from the scaling
regime to dark energy. To this effect, the massive neutrino matter seems
to be a convenient device to achieve the set goal. The non-minimally
coupled massive neutrino-matter contributes to the effective
potential of the field $\phi$ at late stages and induces a minimum in it, where the
scalar field is trapped
giving rise to dark energy dominated epoch.

The non-minimal coupling crucially affects the neutrino masses.
Neutrinos behave as a massless fluid in most of the cosmic history, and
the masses manifest at late-times when neutrinos turn
non-relativistic. As a result, the constraint on the neutrino masses
from early universe observations is relaxed. By making use of the
CosmoMC package, we have shown that the allowed $\Sigma m_{\nu}$ at
the present time is enhanced to
around 1~eV that is significantly larger than in the standard
$\Lambda$CDM model. We conclude that our model is in excellent
agreement with observations and presents a successful scheme of the
unification of primordial dark energy driving inflation in the very early Universe and present dark energy producing accelerated expansion of the present Universe. We should mention that the
generic feature of quintessential inflation is related to the
presence of the kinetic regime after inflation, which gives rise to blue
spectrum of relic gravity waves at very small scales (see Ref. \cite{safia} for details).

As we pointed out, there is a significant enhancement of the allowed
$\Sigma m_{\nu}$ in our model, which could provide room for a sterile neutrino.
In this case, the sterile neutrino with mass around 1 eV, as suggested by anomalies in 
neutrino oscillation experiments, is possible.
If such a possibility is experimentally confirmed, it might
signal the demise of $\Lambda$CDM, since it is well known that
incorporating the neutrinos masses leads to suppression of
matter power spectrum in the framework of $\Lambda$CDM cosmology or slowly rolling 
quintessence. Interestingly, the generic schemes of large-scale modifications of gravity give rise to 
the opposite effect. Indeed, in these schemes, in an approximation valid in the domain of
interest, the modification is captured  replacing $G$ by $G_{eff}$,
which gives enhancement in the matter power spectrum. Hence, this feature could
also allow us to accommodate higher mass neutrinos and salvage
sterile neutrino cosmology~\cite{Motohashi:2012wc,Chudaykin:2014oia}. It would be interesting to
address the issue in the framework of an effective theory
including the most general scalar field action \'a la
Hordeski which leads to second order equations for $\phi$.
%\footnote{A generic large-scale modification of gravity in principle
%constitutes of Einstein gravity along with extra degrees of
%freedom non-minimally coupled to matter sources. $f(R)$ theories
%include one scalar degree of freedom which is massive. In general,
%the extra degrees of freedom may also be massless as is the case of
%the Brans-Dicke gravity. Local gravity constraints can be respected
%using the chameleon mechanism for massive and the Vainshtein screening
%for massless degrees of freedom respectively. The most general case of ghost-free
%scalar degree of freedom is provided by the Hordeski system,
%a generalization of Galileon scalar field cosmology  which naturally occurs in
%dRGT in the decoupling limit.}. 
Such an investigation is left for a future project.

\section*{ACKNOWLEDGMENTS}

The work was supported in part by National Center for Theoretical Science, National Science Council (NSC-101-2112-M-007-006-MY3), 
MoST (MoST-104-2112-M-007-003-MY3 and MoST-106-2917-I-564-055), National Tsing Hua University (104N2724E1) and the Newton International Fellowship (NF160058) from the Royal Society (UK).
A.A.S. was partially supported by the grant RFBR 17-02-01008 and by the Scientific Programme P-7 (sub-programme 7B) 
of the Presidium of the Russian Academy of Sciences.

\end{document}